\documentclass[twocolumn,showkeys,aps,prb,showpacs]{revtex4-1}
\usepackage{graphicx}
\usepackage[CJKbookmarks,dvipdfm,colorlinks,linkcolor=blue,citecolor=blue]{hyperref}

\begin{document}

\title{Monolayer enhanced thermoelectric properties compared with bulk for BiTeBr}

\author{San-Dong Guo and Hui-Chao Li}
\affiliation{School of Physics, China University of Mining and
Technology, Xuzhou 221116, Jiangsu, China}
\begin{abstract}
It is believed that nanostructuring  is an effective way to achieve excellent thermoelectric performance. In the work, by  combining the first-principles calculations and semiclassical Boltzmann transport theory, we investigate the thermoelectric properties of bulk and monolayer BiTeBr  including  both the electron and phonon transports. The generalized gradient approximation (GGA) plus spin-orbit coupling (SOC) is employed  for the electron part,  and GGA for the phonon part. It is found that SOC has important effects on  electronic   transport coefficients because of SOC-induced obvious influences on energy band structures.  In p-type doping, monolayer has larger Seebeck coefficient than bulk in wide doping range, which is beneficial to excellent thermoelectric performance. The calculated average lattice thermal conductivity of bulk  is 1.71   $\mathrm{W m^{-1} K^{-1}}$  at room temperature, which is  close to experimental value 1.3  $\mathrm{W m^{-1} K^{-1}}$.
Calculated results show that monolayer has  better $ZT_e$ and lower lattice thermal conductivity than bulk, which  suggests that monolayer has better thermoelectric performance than bulk. The lower lattice thermal conductivity in monolayer than bulk is due to shorter phonon lifetimes.
By comparing the experimental  electrical conductivity of bulk  with calculated value, the scattering time is determined for 3.3 $\times$ $10^{-14}$ s. Based on electron and phonon transport coefficients, the thermoelectric figure of merit $ZT$  of bulk and monolayer are calculated. It is found that monolayer has higher peak $ZT$ than bulk, and the peak $ZT$  of monolayer can be as high as 0.55 in n-type doping and 0.75 in p-type doping at room temperature. These results  imply that monolayer BiTeBr may be a potential  two-dimensional (2D) thermoelectric material, which can  stimulate further experimental works to synthesize monolayer BiTeBr.

\end{abstract}
\keywords{Bulk and monolayer;  Power factor; Thermal conductivity}

\pacs{72.15.Jf, 71.20.-b, 71.70.Ej, 79.10.-n}

\maketitle

\section{Introduction}
Thermoelectric materials are of interest due to  potential applications in energy conversion devices, and  make  essential contributions to the crisis of energy\cite{s1,s2}. The dimensionless figure of merit $ZT$, defined as $ZT=S^2\sigma T/(\kappa_e+\kappa_L)$, can describe the performance of thermoelectric materials, where S, $\sigma$, T, $\kappa_e$ and $\kappa_L$ are the Seebeck coefficient, electrical conductivity, working temperature, the electronic and lattice thermal conductivities, respectively.  According to expression of $ZT$, a potential thermoelectric  material requires high power factor ($S^2\sigma$) and low thermal conductivity ($\kappa=\kappa_e+\kappa_L$).  Unfortunately, Seebeck coefficient and   electrical conductivity  are oppositely  proportional to carrier concentration.  Therefore, searching for high-performance thermoelectric materials is interesting and challenging.

To improve  $ZT$ of bulk materials,  many strategies have been proposed, such as bands convergence to enhance the Seebeck coefficient by strain or doping\cite{s1,s3,t4,s5} and  phonon  engineering to reduce lattice thermal conductivity by alloying or introducing layered structures\cite{s6,s7}.  To improve $ZT$,  nanostructuring  is another effective way, which is firstly proposed  by Hicks and Dresselhaus in 1993\cite{s8,s9}. The low-dimensional materials can
dramatically improve the power factor due to the sharp
peaks in the electronic density of states (DOS) near the Fermi
energy,  which  can induce asymmetry
between holes and electrons transport and  enhance  electrical conductivity,  and then can produce  large Seebeck coefficient.
Great progress by nanostructuring to enhance $ZT$ has been made, such as  $\mathrm{Bi_2Te_3/Sb_2Te_3}$\cite{s10}, $\mathrm{PbTe/Ag_2Te}$\cite{s11}  and silicon nanowires \cite{s12}.
\begin{figure}
  \includegraphics[width=5.0cm]{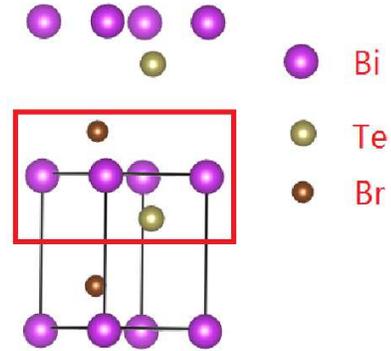}
  \caption{(Color online) The  crystal structures of BiTeBr: the frame surrounded by a black box is bulk unit cell, and the monolayer is presented by a red box. }\label{struc}
\end{figure}

Due to noncentrosymmetric crystal structure and strong SOC, bulk BiTeX (X = Cl, Br, I)  exhibit a giant Rashba-type spin splitting\cite{s13,s14}. Both in theory and in experiment, it is found  that  pressure can produce   a topological transition in BiTeX (X =  Br, I)\cite{q10,q101,q102,q103}.
The thermoelectric properties of  BiTeI  have been investigated  both theoretically and experimentally\cite{q11,q111,q13}, and the  thermoelectric performance  can be enhanced  in Cu-intercalated BiTeI \cite{q131}, through Br-substitution\cite{q132} and by pressure\cite{q133}.  Experimentally, the thermoelectric properties of  BiTeBr have also been studied, whose thermoelectric efficiency  is better than that of BiTeI\cite{q11,q111}.
Theoretically,  monolayers BiTeX (X = Br, I) have been predicted based on the first-principles calculations, which can also produce a giant Rashba spin splitting\cite{q134,q1340}. Recently, the thermoelectric properties of 2D materials, such as semiconducting transition-metal dichalcogenide monolayers, orthorhombic group IV-VI monolayers and group-VA elements (As, Sb, Bi) monolayers\cite{t1,t2,t3,t4,t5,t500}, have been widely investigated.

\begin{figure}
  \includegraphics[width=8cm]{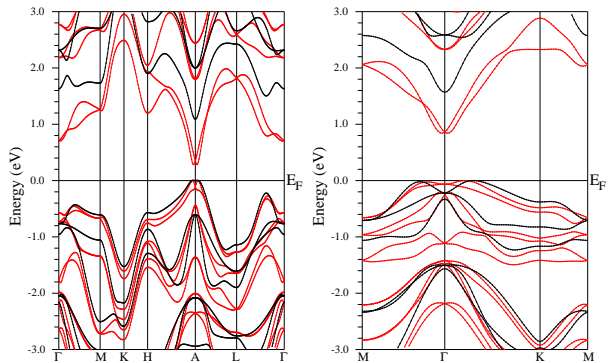}
  \caption{(Color online) The energy band structures of bulk (Left) and monolayer (Right) BiTeBr  using GGA (Black lines)  and GGA+SOC (Red lines).}\label{band}
\end{figure}

 \begin{table}[!htb]
\centering \caption{The  lattice constants $a$ and $c$ ($\mathrm{{\AA}}$); the calculated energy band gaps  using GGA $G$ (eV) and GGA+SOC $G_{so}$ (eV); $G$-$G_{so}$ (eV);  Rashba energy $E_{R}$ (meV). }\label{tab}
  \begin{tabular*}{0.48\textwidth}{@{\extracolsep{\fill}}cccccccc}
  \hline\hline
Name& $a$ & $c$ &$G$& $G_{so}$&$G$-$G_{so}$& $E_{R}$\\\hline\hline
Bulk&4.27&6.46&1.09&0.28&0.81 &63\\\hline
Monolayer&4.37&-&1.57&0.84&0.73&19\\\hline\hline
\end{tabular*}
\end{table}
Here,  we investigate  thermoelectric properties of bulk and monolayer  BiTeBr by  the first-principle calculations and Boltzmann transport theory. It is found that SOC has important effects on electronic   transport coefficients, which has also been found in bulk BiTeI\cite{q133}.  The predicted  average room-temperature lattice thermal conductivity for bulk BiTeBr  is 1.71   $\mathrm{W m^{-1} K^{-1}}$, being close to experimental value 1.3  $\mathrm{W m^{-1} K^{-1}}$\cite{q11,q111}. The room-temperature lattice thermal conductivity for monolayer BiTeBr is lower than one of bulk due to shorter phonon lifetimes.
The scattering time  can be attained by the comparison between experimental and theoretical electrical conductivity. Finally, the thermoelectric figure of merit $ZT$ of bulk and monolayer  are calculated. It is found that monolayer has more higher peak value of $ZT$ than bulk  due to better $ZT_e$ and lower lattice thermal conductivity, which shows monolayer can improve thermoelectric performance compared with bulk for BiTeBr.

The rest of the paper is organized as follows. In the next section, we shall
describe computational details about the first-principle and transport coefficients calculations. In the third section, we shall present the electronic structures and  thermoelectric properties of bulk and monolayer BiTeBr. Finally, we shall give our discussions and conclusion in the fourth section.

\section{Computational detail}
 A full-potential linearized augmented-plane-waves method
within the density functional theory (DFT) \cite{1} is employed to study electronic structures of bulk and monolayer BiTeBr, as implemented in the package WIEN2k \cite{2}. The  free  atomic position parameters  are optimized using GGA of Perdew, Burke and  Ernzerhof  (GGA-PBE)\cite{pbe} with a force standard of 2 mRy/a.u..
The SOC is included self-consistently \cite{10,11,12,so} due to strong  Rashba
spin splitting, giving rise to important effects on electronic transport coefficients. The convergence results are determined
by using  5000 k-points in the
first Brillouin zone (BZ) for the self-consistent calculation, making harmonic expansion up to $\mathrm{l_{max} =10}$ in each of the atomic spheres, and setting $\mathrm{R_{mt}*k_{max} = 8}$ for the plane-wave cut-off.

\begin{figure*}
  \includegraphics[width=15cm]{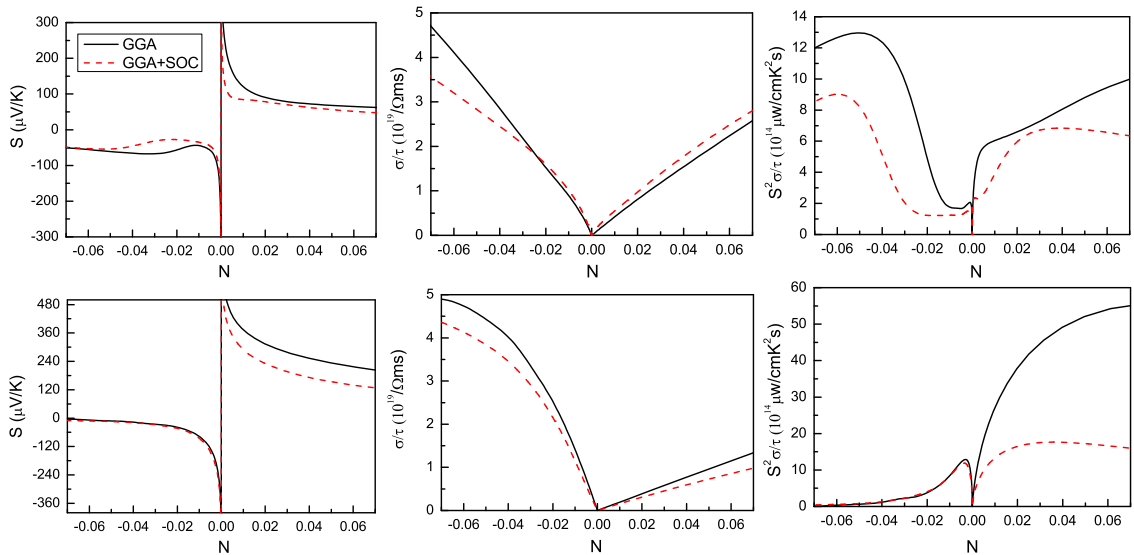}
  \caption{(Color online) At room temperature (300 K),  transport coefficients of bulk (Top) and monolayer (Bottom) BiTeBr as a function of doping level (N), including  Seebeck coefficient S,  electrical conductivity with respect to scattering time  $\mathrm{\sigma/\tau}$ and power factor with respect to scattering time $\mathrm{S^2\sigma/\tau}$ calculated with GGA  and GGA+SOC. }\label{s0}
\end{figure*}
\begin{figure}
  \includegraphics[width=8cm]{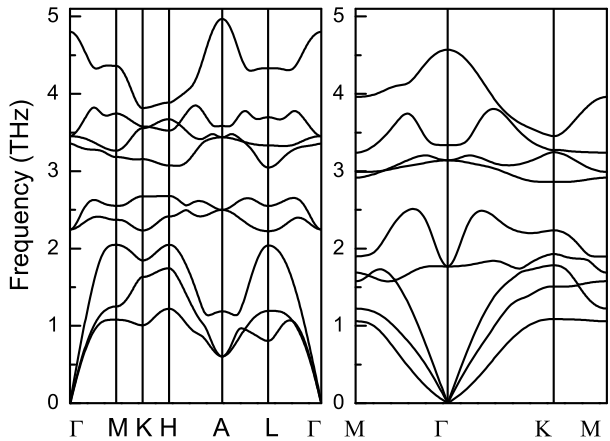}
  \caption{Phonon band structures of bulk (Left) and monolayer (Right) BiTeBr using GGA-PBE.}\label{ph}
\end{figure}

 Based on the results of electronic
structure, transport coefficients for electron part
are calculated through solving Boltzmann
transport equations within the constant
scattering time approximation (CSTA),  as implemented in
BoltzTrap\cite{b}, which shows reliable results in many classic thermoelectric
materials\cite{b1,b2,b3}. To
obtain accurate transport coefficients, we set the parameter LPFAC for 10 in bulk and for 20 in monolayer, and use 4524 (5250) k-points  in the  irreducible BZ of bulk (monolayer) for the energy band calculation.

The  lattice thermal conductivity is performed
by using Phono3py+VASP codes\cite{pv1,pv2,pv3,pv4}.
 The all-electron projector augmented
wave method\cite{pv3} is adopted, and the structures of bulk (monolayer) BiTeBr are relaxed until the atomic forces are less than $10^{-4}$ eV/ $\mathrm{{\AA}}$, using a
20 $\times$ 20 $\times$ 16 (20 $\times$ 20 $\times$ 6) k-point meshes,  with a  kinetic energy cutoff of 400 eV.
The electronic stopping criterion is $10^{-8}$ eV.
The  lattice thermal conductivities of bulk and monolayer BiTeBr are carried out with the single mode relaxation time approximation (RTA) and linearized phonon Boltzmann equation  using Phono3py code\cite{pv4}. The interatomic force constants (IFCs) are calculated by
the finite displacement method.
The second-order harmonic IFCs of bulk (monolayer) BiTeBr
are performed using a 3 $\times$ 3 $\times$ 2 (5 $\times$ 5 $\times$ 1)  supercell  containing
54 (75) atoms with k-point meshes of 3 $\times$ 3 $\times$ 2 (2 $\times$ 2 $\times$ 1). The phonon dispersions of bulk and monolayer BiTeBr can be attained by harmonic IFCs, as implemented in the Phonopy package\cite{pv5}.  The third-order anharmonic IFCs of bulk (monolayer) BiTeBr are calculated using a 3 $\times$ 3 $\times$ 2 (3 $\times$ 3 $\times$ 1)
supercells containing 54 (27) atoms with k-point meshes of 3 $\times$ 3 $\times$ 2 (6 $\times$ 6 $\times$ 1), and the total number of displacements
is 1413 (711). To compute lattice thermal conductivities, the
reciprocal spaces of the primitive cells  are sampled using a 20 $\times$ 20 $\times$ 16 (40 $\times$ 40 $\times$ 2)  meshes.

For 2D material, the calculated lattice  thermal conductivity  depends on the length of unit cell used in the calculations along z direction\cite{2dl}, which should be normalized by multiplying $Lz/d$, where $Lz$ is the length of unit cell along z direction  and $d$ is the thickness of 2D material, but the $d$  is not well defined. The electrical conductivity and electronic thermal conductivity
are the same with lattice  thermal conductivity, which should also be normalized by multiplying $Lz/d$. However, The dimensionless figure of merit $ZT$ is independent of  the length of unit cell used in the calculations along z direction.

\section{MAIN CALCULATED RESULTS AND ANALYSIS}
Bulk BiTeBr has a layered structure along  c axis with  space group  being $P3m1$, whose unit cell contains three atoms with Bi atom sandwiched between  Te and  Br atoms, forming a triple layer.  Because of weak van der Waals interactions between the adjacent triple layers, monolayer BiTeBr can be exfoliated in experiment. The  schematic crystal structures of bulk and monolayer BiTeBr are presented in \autoref{struc}. In our calculations, the experimental values of  lattice constants of bulk BiTeBr (a=b=4.27 $\mathrm{\AA}$, c=6.46 $\mathrm{\AA}$) are used\cite{lc}.  The  lattice constants of monolayer BiTeBr are optimized  within GGA-PBE, and the optimized value is a=b=4.37 $\mathrm{\AA}$, which is less than one of monolayer BiTeI(4.42 $\mathrm{\AA}$)\cite{q1340}.
The unit cell  of  monolayer BiTeBr, containing one Bi, one  Te and one Br atoms,  is constructed with the vacuum region of larger than 15 $\mathrm{{\AA}}$ to avoid spurious interaction.  All the free atomic positions of both bulk and monolayer  are optimized within GGA-PBE.

The energy band structures of bulk and monolayer BiTeBr using GGA and GGA+SOC are shown in \autoref{band}.
The conduction bands of both bulk and monolayer  are mainly composed of the 6p-states  of Bi, while the valence bands
are dominated by  Te-p and Br-p states. The SOC can lead to a huge reduction in the band gap. The GGA and GGA+SOC gaps of bulk are 1.09 eV and 0.28 eV, respectively, while ones of monolayer  are 1.57 eV and 0.84 eV.  The GGA+SOC gap of bulk agrees well with other theoretical value 0.31 eV\cite{the1}.
It is clearly seen that monolayer  has larger gap than  bulk, which is similar to semiconducting transition-metal dichalcogenide compounds. The outlines of band
structures are strongly modified due to spin-orbit splitting.
The splitting makes  the conduction band minimum (CBM) of the bulk  (monolayer) BiTeBr
deviate slightly from high symmetry A ($\Gamma$) point,  forming a significant Rashba spin splitting.
The Rashba energy ($E_R$) of bulk (monolayer) BiTeBr, which is defined as the energy between the CBM and the band crossing point of conduction bands at high symmetry point A  ($\Gamma$), is 63 meV  (19 meV) . The bulk Rashba energy is in agreement with previous calculated value 55 meV\cite{the1}.

Next, the  electronic   transport coefficients  are performed  using CSTA Boltzmann theory,  Seebeck coefficient of which  is independent of scattering time. The doping level, which is defined as  electrons (minus value) or holes (positive value) per unit cell,  can be simulated   by simply shifting  Fermi level into conduction (n-type doping) or valence (p-type doping) bands within the framework of  rigid band approach.
The n-type doping  produces   the negative Seebeck coefficient, while  the p-type doping leads to  the positive Seebeck coefficient. At room temperature , Seebeck coefficient S,  electrical conductivity with respect to scattering time  $\mathrm{\sigma/\tau}$ and power factor with respect to scattering time $\mathrm{S^2\sigma/\tau}$  of bulk and monolayer BiTeBr as a function of doping level using GGA and GGA+SOC are shown  in \autoref{s0}. Note: The electrical conductivity of monolayer has been  normalized by multiplying $Lz/d$, where $d$ is the lattice constant $c$ of bulk. For bulk BiTeBr, SOC has a reduced effect on Seebeck coefficient, which leads to SOC-reduced power factor. For monolayer BiTeBr, SOC has a slight effect on n-type  Seebeck coefficient, but has a obviously reduced effect on p-type Seebeck coefficient. The SOC influence on power factor of monolayer  is the same with one on Seebeck coefficient. It is found that monolayer  has larger p-type Seebeck coefficient than bulk, leading to larger power factor, if the scattering time  $\mathrm{\tau}$ is assumed to be the same.
The p-type Seebeck coefficient of bulk is larger than 200 $\mu$V/K with doping level being less than 5.9$\times$$10^{-4}$, while monolayer one with doping level  being less than 2.9$\times$$10^{-2}$  is larger than 200 $\mu$V/K. That shows monolayer has more wider doping range than bulk for efficient thermoelectric application.

\begin{figure}
  \includegraphics[width=7cm]{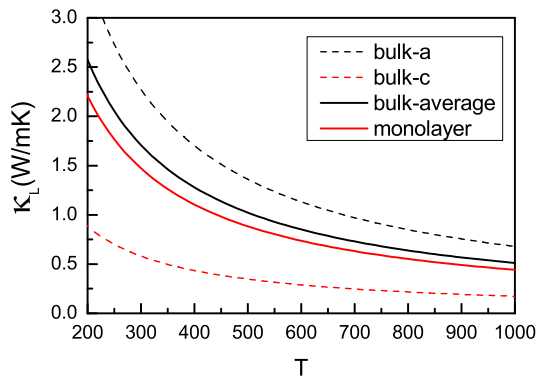}
  \caption{(Color online) The lattice thermal conductivities of  bulk  and monolayer  BiTeBr   using GGA-PBE.}\label{kl}
\end{figure}

\begin{figure}
  \includegraphics[width=7.0cm]{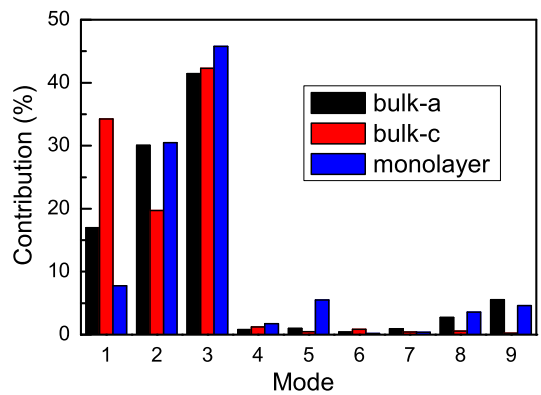}
  \caption{(Color online) The phonon modes contributions toward total lattice thermal conductivity (300 K). 1, 2, 3 represent acoustic branches and 4, 5, 6, 7, 8, 9 for optical branches.}\label{mkl}
\end{figure}
\begin{figure*}
  \includegraphics[width=12.0cm]{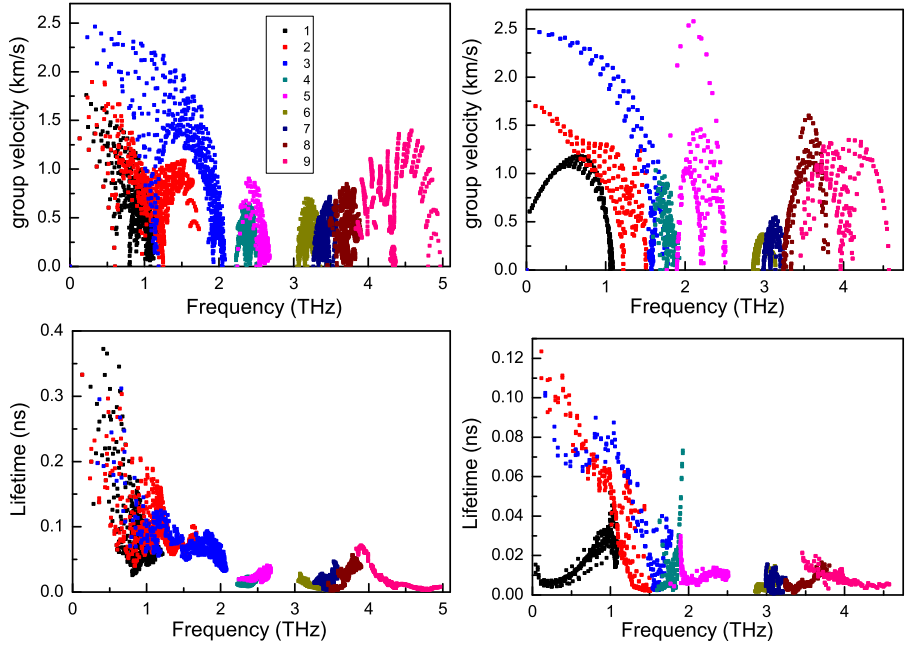}
  \caption{(Color online) The phonon mode group velocities (Top) and room-temperature phonon lifetimes (Bottom)  of  bulk (Left) and monolayer (Right)  BiTeBr  in the first BZ. 1, 2, 3 represent acoustic branches and 4, 5, 6, 7, 8, 9 for optical branches. }\label{mkl1}
\end{figure*}

Based on the harmonic IFCs,  phonon band structures of bulk and monolayer BiTeBr are calculated,
 which are shown in \autoref{ph} along  high-symmetry pathes. Due to three atoms per unit cell, the phonon dispersions of bulk and monolayer BiTeBr contain  3 acoustic and 6 optical phonon branches.
 It is found that three  acoustic branches of bulk  are linear near the $\Gamma$ point. However, the longitudinal acoustic (LA) and transverse acoustic (TA)
branches  are linear near the $\Gamma$ point, while the z-direction acoustic (ZA) branch is quadratic.
Calculated results also show that the whole  branches of monolayer  move toward lower energy compared to ones of bulk.
There is a phonon band gap of 0.19 THz between acoustic and optical branches for bulk, but ZO branch crosses with the  LA branch for monolayer, which can lead to more scattering channel.

 The lattice  thermal conductivities of bulk and monolayer BiTeBr are calculated within  the linearized phonon Boltzmann equation,
 and the  lattice  thermal conductivity is assumed to be independent of  doping level, which is reasonable in many thermoelectric materials\cite{lc11,lc21}.
 Because of  crystal symmetry of bulk, the lattice  thermal conductivities along a and b axises (the in-plane  direction) are equivalent, but they are  different from one along c axis (the cross-plane direction). Therefore, the  lattice  thermal conductivities along a and c axises and the average one ($\kappa_L(av)$=($\kappa_L(xx)$+$\kappa_L(yy)$+$\kappa_L(zz)$)/3) as a function of temperature are plotted in \autoref{kl}.
Calculated results show that the lattice thermal conductivity
 exhibits  obvious anisotropy, and  the lattice thermal conductivity
along c axis is lower than that along  a axis. The  corresponding lattice thermal conductivity  at 300 K is 2.27 $\mathrm{W m^{-1} K^{-1}}$ along a axis and 0.58 $\mathrm{W m^{-1} K^{-1}}$ along c axis, respectively. The average room-temperature lattice thermal conductivity is 1.71 $\mathrm{W m^{-1} K^{-1}}$, which is larger than experimental value 1.3 $\mathrm{W m^{-1} K^{-1}}$\cite{q11,q111}. This difference may be due to defect in experimental sample, which leads to lower
lattice thermal conductivity. The  lattice  thermal conductivity of monolayer  along a axis  ($\kappa_L(xx)$=$\kappa_L(yy)$=$\kappa_L(av)$) as a function of temperature is also plotted in \autoref{kl} with the  thickness $d$ being lattice constant $c$ of bulk. The room temperature lattice thermal conductivity of monolayer  is 1.47 $\mathrm{W m^{-1} K^{-1}}$, which is lower than average one 1.71 $\mathrm{W m^{-1} K^{-1}}$ of bulk.
The  phonon modes contributions of bulk (the in-plane and cross-plane directions) and monolayer BiTeBr to the total lattice
thermal conductivity at 300K are plotted in \autoref{mkl}.
 It is found that the acoustic phonon branches dominate  lattice thermal conductivity, and the acoustic branches  comprise around 88.46\% for bulk along a direction, 96.25\% for bulk along c direction and  84.00\% for monolayer, respectively.
Along the in-plane direction for bulk, the  contribution from the first (17.00\%)  acoustic branch   is the smallest  in acoustic branches, and  the eighth (2.75\%)  and ninth (5.56\%)  optical branches have obvious contributions. However, the second  acoustic branch (19.71\%)  provides the smallest   contribution in acoustic branches along the cross-plane direction for bulk.
For monolayer, the ZA branch (7.77\%)  provides the smallest   contribution in acoustic branches, and  the fifth (5.50\%) , eighth (3.59\%)  and ninth (4.61\%)  optical branches have relatively large contributions.

To understand deeply phonon transports  of bulk and monolayer BiTeBr,  the mode level phonon group velocities
and lifetimes are plotted in \autoref{mkl1}.  It is clearly seen that group velocities of bulk have the same  order of magnitude with ones of monolayer, but
the phonon lifetimes of monolayer are  very shorter than ones of bulk, which leads to lower lattice thermal conductivity for monolayer than bulk.
For both bulk and monolayer, the optical branches have relatively large group velocities, which leads to  relatively large contributions from optical branches to total lattice  thermal conductivity. Moreover, for the fifth optical branch of monolayer,  the maximum  group velocity   is 2.58 $\mathrm{km s^{-1}}$, which is larger than maximum  group velocity of acoustic branches. This leads to  the largest  contribution from the fifth optical branch in optical  branches, which is close to one of ZA branch. The most of  phonon lifetimes of ZA branch for  monolayer are very shorter than ones of LA and TA branches, which produces very little contributions  to total lattice  thermal conductivity.
\begin{figure*}[htp]
    \includegraphics[width=12.0cm]{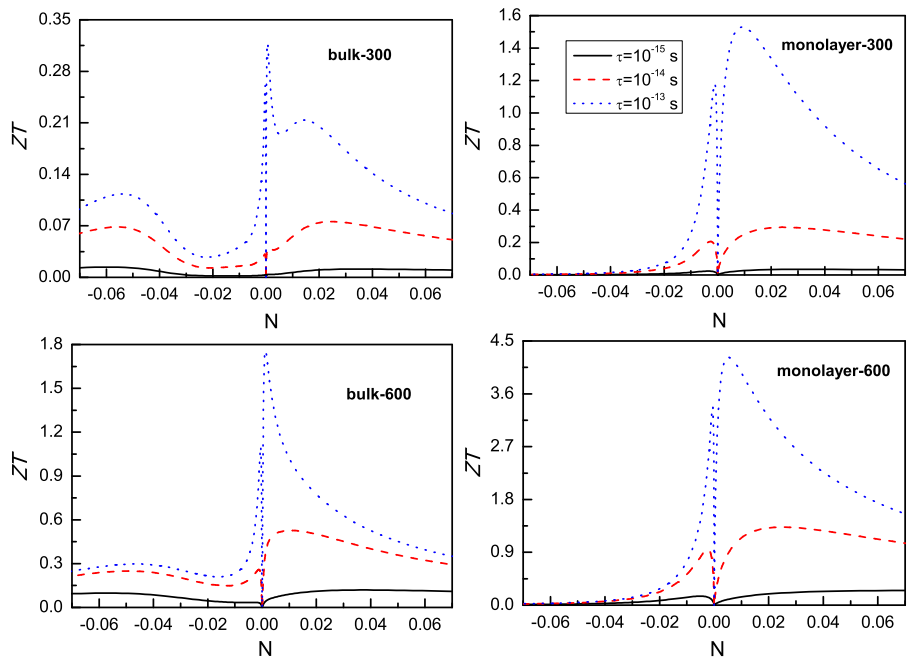}
  \caption{(Color online) The  $ZT$ of bulk  and monolayer  BiTeBr as a function of doping level with temperature  being 300 and 600 K by using three different scattering time $\mathrm{\tau}$  (1 $\times$ $10^{-15}$ s, 1 $\times$ $10^{-14}$ s and 1 $\times$ $10^{-13}$ s).}\label{szt}
\end{figure*}

 Based on the power factor and electronic thermal conductivity, monolayer has more higher $ZT_e$ than bulk, where $ZT_e$ is defined as $S^2\sigma T/\kappa_e$. The $ZT$ is connected to $ZT_e$ by the relation $ZT$=$ZT_e$$\times$$\kappa_e/(\kappa_e+\kappa_L)$. Here, the  $\kappa_e$ is calculated  by the Wiedemann-Franz law with the Lorenz number of 2.4$\times$$10^{-8}$ $\mathrm{W\Omega K^{-2}}$.  To attain the figure of merit $ZT$, only scattering time $\mathrm{\tau}$ is unknown. Here,  a range of reasonable relaxation time is used in our calculations from 1 $\times$ $10^{-15}$ s to 1 $\times$ $10^{-13}$ s.  The  $ZT$ values of bulk  and monolayer  BiTeBr as a function of doping level with temperature  being 300 and  600 K are plotted in \autoref{szt} with the scattering time $\mathrm{\tau}$  equaling  1 $\times$ $10^{-15}$ s, 1 $\times$ $10^{-14}$ s and 1 $\times$ $10^{-13}$ s. As the scattering time $\mathrm{\tau}$ increases, the $ZT$ moves toward its  upper limit $ZT_e$. Calculated results show that monolayer  has larger peak $ZT$ than bulk in both n- and p-type doping  with the scattering time $\mathrm{\tau}$  from 1 $\times$ $10^{-15}$ s to 1 $\times$ $10^{-13}$ s, which shows monolayer enhanced thermoelectric properties compared with bulk. It is found that  the p-type doping has more excellent thermoelectric properties than n-type doping for both bulk and monolayer.

\begin{figure}
  \includegraphics[width=7cm]{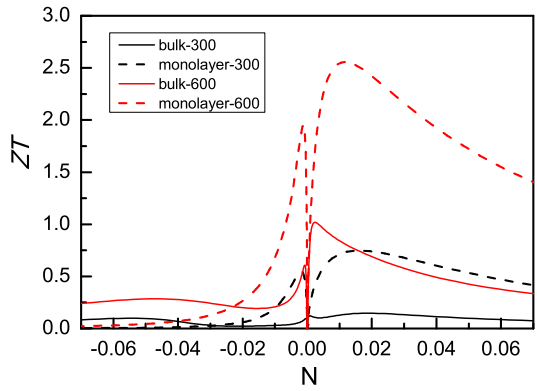}
  \caption{(Color online) The  $ZT$ of bulk  and monolayer  BiTeBr as a function of doping level at 300 and 600 K with the scattering time $\mathrm{\tau}$  being 3.3 $\times$ $10^{-14}$ s.}\label{szt1}
\end{figure}

Due to the complexity of various carrier scattering mechanisms, calculating scattering time $\tau$ from the first principles is difficulty and challenging, but it can be calculated by comparing experimental value  of electronic conductivity with
the calculated value of $\mathrm{\sigma/\tau}$. Experimentally,  the electronic conductivity of bulk BiTeBr at room temperature is about 670 $\mathrm{\Omega^{-1} cm^{-1}}$ with n-type doping concentration $\mathrm{1.4\times10^{19} cm^{-3}}$\cite{q11,q111}.
 The scattering time $\tau$ is found to be  3.3 $\times$ $10^{-14}$ s. The calculated Seebeck coefficient with n-type doping concentration $\mathrm{1.4\times10^{19} cm^{-3}}$  at 300 K is -89.5  $\mu$V/K, which is close to experimental value -115 $\mu$V/K\cite{q11,q111}. The  electronic thermal conductivity $\mathrm{\kappa_e}$ with n-type doping concentration $\mathrm{1.4\times10^{19} cm^{-3}}$ is calculated  using $\tau$=3.3 $\times$ $10^{-14}$ s, and the calculated valve is 0.52  $\mathrm{W m^{-1} K^{-1}}$, which is  in good agreement with experimental value 0.5  $\mathrm{W m^{-1} K^{-1}}$\cite{q11,q111}.  The scattering time $\tau$ attained from bulk BiTeBr  is also used in monolayer BiTeBr, and the  $ZT$ values of bulk  and monolayer  as a function of doping level at 300 and 600 K  are plotted in \autoref{szt1}.
 It is found that the peak $ZT$  values of monolayer  in both n- and p-type doping are larger than ones of bulk.  The  n- and p-type peak $ZT$ of monolayer  is 0.55 and 0.75 at 300 K, and  1.95 and 2.56 at 600 K. These results  imply that monolayer BiTeBr may be a potential  two-dimensional (2D) thermoelectric material.

\section{Discussions and Conclusion}
The SOC  can produce a giant effect on energy band structures near the Fermi level by  removing  band degeneracy and modifying the outline of bands, which can lead to remarkable reduced effects
on the Seebeck coefficient, and further can induce detrimental power factor.
Similar  detrimental effects on power factor  are also found in  $\mathrm{Mg_2X}$
(X = Si, Ge, Sn)\cite{so1},  half-Heusler ANiB (A = Ti, Hf,
Sc, Y; B = Sn, Sb, Bi)\cite{gsd1} and semiconducting transition-metal dichalcogenide monolayers $\mathrm{MX_2}$ (M=Zr, Hf, Mo and Pt; X=S, Se and Te)\cite{gsd20,t4,gsd3}. However, the SOC  also can lead to
observably enhanced power factor in monolayers $\mathrm{WX_2}$ (X=S, Se and Te) due to the
 bands converge induced by SOC\cite{gsd3}. The maximum power factors (MPF) in unit of $\tau\times10^{14}$$\mathrm{\mu W/(cm K^2 s)}$ of  bulk in n-type doping and monolayer in p-type doping  are   extracted  with GGA and GGA+SOC at 300K, and the corresponding GGA and GGA+SOC values are 12.96 and 9.00 for bulk,  55.74 and 17.65 for monolayer. The MPF  with SOC is predicted to be  about 30.56\% (bulk) and 68.34\% (monolayer) smaller than that without SOC.  So, it is very important  for electronic   transport coefficients of BiTeBr  to include SOC.

Strain or pressure has been proved to be very effective to achieve enhanced thermoelectric properties in both bulk and 2D materials. Low-dimensional electronic structures  can occur  in Rashba semiconductor BiTeI\cite{q13}, and pressure can lead to two-dimensional-like DOS in the conduction  bands by tuning  Rashba spin-splitting, which can  induce significantly enhanced power factor in n-type doping by pressure\cite{q133}. The bulk BiTeBr has similar energy band structures with bulk BiTeI, so it is possible to achieve improved power factor in bulk BiTeBr by pressure. Strain-enhanced  power factor  is observed in monolayer $\mathrm{MoS_2}$\cite{gsd20}, $\mathrm{PtSe_2}$\cite{t4} and $\mathrm{ZrS_2}$\cite{e4-1} due to  bands converge induced by strain. It has been proved that  the Rashba spin-splitting can be  modified significantly by the biaxial strain in monolayer BiTeBr\cite{q134}, and it is  possible to achieve enhanced power factor in monolayer BiTeBr by tuning  Rashba spin-splitting.

The thermoelectric properties of many 2D materials have been investigated\cite{t1,t2,t3,t4,t5,t500}, and low lattice thermal conductivity is very important to achieve potential thermoelectric materials.  To compare the lattice thermal conductivities of various 2D materials,  the same thickness should be used. Here, the same thickness of 3.35 $\mathrm{\AA}$ is used\cite{2dl}, and the lattice thermal conductivity of monolayer BiTeBr is 2.84 $\mathrm{W m^{-1} K^{-1}}$, which is lower than that of orthorhombic group IV-VI monolayers (5.24$\sim$15.80 $\mathrm{W m^{-1} K^{-1}}$)\cite{2dl},  semiconducting transition-metal dichalcogenide monolayers (18.55$\sim$261.0 $\mathrm{W m^{-1} K^{-1}}$)\cite{2dl} and group-VA elements (As, Sb, Bi) monolayers (4.78$\sim$48.09 $\mathrm{W m^{-1} K^{-1}}$)\cite{t500}. Therefore, monolayer BiTeBr may be a potential 2D thermoelectric material due to lower lattice thermal conductivity compared to other  well-studied 2D materials.

In summary,  based mainly on the reliable first-principle calculations, the thermoelectric properties  of  bulk and monolayer BiTeBr are investigated. The electron part is performed using GGA+SOC, and  phonon part is calculated using GGA.
Calculated results show that SOC has important effects on electronic   transport coefficients due to obvious SOC influences on electronic structures.
The calculated  average bulk  lattice thermal conductivity at room temperature  is  1.71 $\mathrm{W m^{-1} K^{-1}}$, being close to experimental value 1.3 $\mathrm{W m^{-1} K^{-1}}$.  The  monolayer lattice thermal conductivity (300 K) is 1.47  $\mathrm{W m^{-1} K^{-1}}$ with $d$ being $c$ of bulk, which is lower than that of bulk. The scattering time $\tau$ is determined by fitting the calculated
electronic  conductivity with experimental measurement, and the attained scattering time $\tau$ is 3.3 $\times$ $10^{-14}$ s.
The monolayer  has larger  peak $ZT$  in both n- and p-type doping than bulk due to higher $ZT_e$ and  lower lattice thermal conductivity.   The present
work can  encourage further  experimental efforts to achieve  monolayer BiTeBr, and  then to investigate it's thermoelectric performance.

\begin{acknowledgments}
This work is supported by the National Natural Science Foundation of China (Grant No.11404391 and  Grant No.11404392) and the Fundamental Research Funds for the Central Universities (Grant No.2014QNA53).  We are grateful to the Advanced Analysis and Computation Center of CUMT for the award of CPU hours to accomplish this work.
\end{acknowledgments}

\end{document}